\documentclass{PoS}

\title{Baryon interactions from lattice QCD with physical masses -- $S=-2$ sector --}

\ShortTitle{Baryon interactions from lattice QCD with physical masses -- $S=-2$ sector --}

\author{{\speaker{Kenji Sasaki}${^{a}}$, Sinya~Aoki${^{abc}}$, Takumi~Doi${^{b}}$, Shinya~Gongyo${^{bd}}$, Tetsuo~Hatsuda${^{be}}$, Yoichi~Ikeda${^{bf}}$,
Takashi~Inoue${^{bg}}$ Takumi~Iritani${^{b}}$, Noriyoshi~Ishii${^{bf}}$, Takaya~Miyamoto${^{ab}}$ and
Keiko~Murano${^{bf}}$}\\ 
${^a}$ Yukawa Institute for Theoretical Physics, Kyoto University, Kyoto 606-8502, Japan \\
${^b}$ Theoretical Research Division, Nishina Center, RIKEN, Wako 351-0198, Japan \\
${^c}$ Center for Computational Sciences, University of Tsukuba, Ibaraki 305-8571, Japan \\
${^d}$ CNRS, Laboratoire de Math\'ematiques et Physique Th\'eorique, Universit\'ede Tours, 37200 France \\
${^e}$ iTHEMS Program and iTHES Research Group, RIKEN, Wako 351-0198, Japan \\
${^f}$ Research Center for Nuclear Physics (RCNP), Osaka University, Osaka 567-0047, Japan \\
${^g}$ Nihon University, College of Bioresource Sciences, Kanagawa 252-0880, Japan \\
        E-mail: \email{kenjis@yukawa.kyoto-u.ac.jp}}

\abstract{
The strangeness $S=-2$ baryon-baryon interaction is investigated directly from the fundamental theory of the strong interaction, QCD.
The HAL QCD method enables us to extract baryon interactions from the Nambu-Bethe-Salpeter wave functions without using any experimental information.
We present our latest result on the $S = -2$ baryon interactions and discuss the H-dibaryon state using potentials which are calculated by using the (almost) physical point gauge configurations with large lattice volume of$(8.1{\rm{fm}})^4$ generated on the K-computer.}

\FullConference{34th annual International Symposium on Lattice Field Theory\\
		24-30 July 2016\\
		University of Southampton, UK}

\begin{document}
\section{Introduction}
\label{intro}
The strangeness $S=-2$ sector of baryon interactions are key to investigate the possibility of exotic states, the structures of hypernuclear states and the properties of deep inside neutron stars.
Model calculations of  baryon interactions have been performed on the basis of the flavor $SU(3)$  symmetry in order to reduce some artificial parameters. 
Since a direct scattering experiment of hyperons is not feasible due to the short lifetimes of hyperons, their scattering data are quite limited and are insufficient to construct realistic baryon interactions.

Recent development of lattice QCD simulations enables us to extract information of baryon interactions non-perturbatively.
The framework which relate the energy $E$ of a two-particle state in a finite box to the elastic scattering phase $\delta(E)$ in the continuum was firstly proposed by M.~L\"uscher~\cite{Luscher:1990ux}.
The relation is derived by using the asymptotic behavior of the two-particle Nambu-Bethe-Salpeter (NBS) wave function $\psi(r)$, if the range of the interaction is sufficiently smaller than the size of the box.
An alternative approach to the hadron interactions from lattice QCD has been proposed \cite{Ishii:2006ec, Aoki:2012tk} and has been extensively developed by the HAL QCD Collaboration \cite{Nemura:2008sp, Aoki:2011gt, Murano:2011nz, Inoue:2011ai, Doi:2011gq, HALQCD:2012aa, Murano:2013xxa, Ikeda:2013vwa, Sasaki:2015ifa}, called the HAL QCD method. 
In the method, we define the energy-independent and non-local potential $U(r, r')$ from $\psi(r)$ which obeys the Schr\"odinger type equation in a finite box.
Using the obtained $U$ which receives only weak finite volume effect, we can simply calculate the scattering phase shifts and bound state spectra in infinite space to compare the results with experimental data.
An obvious advantage of the HAL QCD method is that it can be generalized straightforwardly to the case of inelastic scatterings.

 In this paper, we investigate the $BB$ interaction with the explicit $SU(3)$ breaking on the basis of the coupled channel HAL QCD method developed in our previous works\cite{Aoki:2011gt, Sasaki:2015ifa}.

\section{Coupled channel $BB$ potential}
We start from the normalized four-point correlation function $R$ in channel $c$ defined as
\begin{eqnarray}
 R^c_{\mathcal{I}_d}(\vec r, t) 
 &\equiv& \frac{\langle 0 \mid B_{c_1}(\vec{x} +\vec{r},t) B_{c_2}(\vec{x}, t) \overline{\mathcal{I}}_d (t_0 = 0)\vert 0 \rangle}{ \sqrt{Z_{c_1} Z_{c_2}} \exp[-(m_{c_1}+m_{c_2}) t]}
 =\sum_{n} \psi^{c}_{W_n} (\vec{r}) e^{-\Delta W_n^c t} A_d^{W_n} +\cdots, 
\label{EQ.Rdefine}
\end{eqnarray}
where $B_{c_j}(\vec{x},t)$ is an interpolating operator for octet baryon with a channel index, $c$, and particle index, $j=1,2$, and $\sqrt{Z_{c_j}}$ is the corresponding wave-function renormalization factor. 
$\psi^{c}_{W_n}$ denotes the equal-time NBS wave function with the total energy $W_n$.
An effect of source operator, $\overline{\mathcal{I}}_{d}(0)$, emerges as $A_d^{W_n} = \langle W_n \vert \overline{\mathcal{I}}_{d}(0) \vert 0 \rangle$.
The energy from two baryon state in the channel $c$ is denoted as $\Delta W_n^c = W_n - m_{c_1}-m_{c_2}$  with baryon mass $m_{c_i}$
The ellipses in Eq.(\ref{EQ.Rdefine}) denote inelastic contributions from channels which we are not considering. 

A coupled channel potential can be obtained by using the $R$-correlator via the time-dependent Schr\"{o}dinger-like equation~\cite{HALQCD:2012aa} as
\begin{eqnarray}
\left(D_t^c - H_0^c \right) {R^{c}}_{\mathcal{I}_d}(\vec r, t)  &=& \int d^3 r^\prime {U^c}_e(\vec r, \vec r^\prime) {\Delta^c}_e
{R^e}_{\mathcal{I}_d}(\vec r^\prime, t),
\end{eqnarray}
where $ {H_0}^c = - \frac{\nabla^2}{2\mu^c} $ and ${\Delta^c}_e = \exp[-(m_{e_1}+m_{e_2})t] /  \exp[-(m_{c_1}+m_{c_2})t] $.
The operator $D_t^c$ is corresponding to the kinetic energy part of Schr\"odinger equation, ${{k_i^{c}}^2}/{2\mu^c}$, and described up to the second order of the time-derivative for $R$ function or, equivalently, of $\Delta W$ as
\begin{eqnarray}
D_t^c {R^{c}}_{\mathcal{I}_d}(\vec r, t) 
\simeq
-\frac{\partial}{\partial t} {R^{c}}_{\mathcal{I}_d}(\vec r, t) + \frac{1}{8 \mu} \left[ 1 + 3 \left( \frac{m_{c_1}-m_{c_2}}{m_{c_1}+m_{c_2}} \right)^2 \right] \frac{\partial^2}{\partial t^2} {R^{c}}_{\mathcal{I}_d}(\vec r, t)
\end{eqnarray}
where the asymptotic momentum $k_i^{c}$ in the center-of-mass (CM) frame is defined via the 
relativistic energy as
\begin{eqnarray}
W_n = \sqrt{m_{c_1}^2+(k_n^c)^2} + \sqrt{m_{c_2}^2+(k_n^c)^2}.  
\end{eqnarray}
For each element of the coupled-channel potential matrix, we consider the derivative expansion to manage the non-locality of the potential as
\begin{eqnarray}
  {U^{c_1}}_{c_2}(\vec r,\vec r') 
  = ({{V_{\rm{LO}}}^{c_1}}_{c_2}(\vec r)+ {{V_{\rm{NLO}}}^{c_1}}_{c_2}(\vec r)  + \cdots )\delta(\vec r-\vec r^\prime) 
\end{eqnarray}
where N${^n}$LO term is of $O(\vec{\nabla}^n)$ and its convergence has been confirmed for the $NN$ case~\cite{Murano:2011nz}.
In this paper, we truncate the expansion in the leading order and does not split the tensor potential from the leading order term in spin-triplet states for simplicity.


\section{Lattice setup}
\begin{table}
\caption{Baryon masses in units of MeV.}
\begin{center}
\begin{tabular}{c|cccc}
\hline \hline
particle & $N$ & $\Lambda$ & $\Sigma$ & $\Xi$ \\
\hline
mass [MeV] & 953 $\pm$ 7 & 1123 $\pm$ 3 & 1204 $\pm$ 2 & 1332 $\pm$ 1 \\
\hline \hline
\end{tabular}
\end{center}
\label{TAB.HADRON_MASS}
\end{table}

We employ $N_f = 2+1$ gauge configurations which are generated with the Iwasaki gauge action at $\beta = 1.82$ and nonperturbatively ${\cal{O}}(a)$-improved Wilson quark action with $c_{sw} = 1.11$ on the $L^3 \times T = 96^3 \times 96$ lattice~\cite{Ishikawa:2015rho}.
The hopping parameters for light (ud) and strange quarks are choosen as $(\kappa_{ud},\kappa_s) = (0.126117,0.124790)$ corresponding to $m_\pi \simeq 146$ MeV and $m_K \simeq 525$ MeV with $a^{-1} \simeq 2.33$ GeV ($a \simeq 0.085$fm).
This lattice setup brings about the almost physical point simulation of the $BB$ interaction on the large lattice volume of $(8.1{\rm{fm}})^4$ where a finite volume effects of the $BB$ potential could be neglected. 
We calculate quark propagators in consideration of a zero momentum wall source by imposing Coulomb gauge fixing at $t_0$ with the Dirichlet boundary condition in temporal direction at $|t_{DBC}-t_0| = 48$.
The total statistics used in this report amounts to $414$ configurations $\times$ $4$ rotations $\times$ $48$ wall sources.
The forward and backward propagations of baryon four-point correlator are averaged  and four rotated gauge configurations are used to reduce the statistical errors.
An average over the cubic group is taken for the sink operator to project on the S-wave in the $BB$ wave function. 
The baryon masses measured in this setup are listed in Tab.~\ref{TAB.HADRON_MASS}.
Jackknife prescription with the bin of the size $6$ ($69$ configurations $\times$ $5$ trajectries) is used to estimate the statistical errors.

\section{Results and discussions}
The coupled channel $BB$ potential in the strangeness $S=-2$ sector is calculated by using the NBS wave functions at $t-t_0=11$~\footnote{
Although we confirmed that the $t$-dependence of the calculated potentials is mild, it is desirable to choose the larger $t$ in order to suppress contaminations from unconsidered inelastic states.
}
The channels of the $S=-2$ $BB$ system with the total spin of $J=0$ and $J=1$ are summarized in Table~\ref{TAB:S2channel}.
The other sectors of $BB$ interactions are discussed in~\cite{LatProc2016}.
\begin{table}
\caption{Summary of channels with $S=-2$ $BB$ system and their $SU(3)$ decompositions in the symmetry limit are shown. We use the ${^{2S+1}L_J}$ symbols to describe the two-baryon states. 
For $J=0$ state, the $|1\rangle$, $|8_s\rangle$ and $|27\rangle$ indicate the state of flavor singlet, octet and $27$-plet, respectively.
For $J=1$ state, the $|8_a\rangle$, $|10\rangle$ and $|\bar{10}\rangle$ indicate the state of flavor octet, decuplet and anti-decuplet, respectively.
}
\label{TAB:S2channel}
\begin{center}
\begin{tabular*}{0.7\linewidth}{@{\extracolsep{\fill}}llll}
\hline \hline
\multicolumn{2}{l}{Channel} & {State} & {$SU(3)$ description}  \\
\hline \hline 
${^1S_0}$
 & $I=0$  & 
 {$|\Lambda \Lambda \rangle$} & {$= - \sqrt{\frac{1}{8}} | 1 \rangle - \sqrt{\frac{1}{5}} | 8_s \rangle + \sqrt{\frac{27}{40}} | 27 \rangle$} \\ 
 & & 
 {$|N \Xi \rangle$} & {$=  \phantom{-} \sqrt{\frac{1}{2}} | 1 \rangle + \sqrt{\frac{1}{5}} | 8_s \rangle + \sqrt{\frac{3}{10}} | 27 \rangle$} \\
 & & 
 {$|\Sigma \Sigma \rangle$} & {$=  \phantom{-} \sqrt{\frac{3}{8}} | 1 \rangle - \sqrt{\frac{3}{5}} | 8_s \rangle - \sqrt{\frac{1}{40}} | 27 \rangle$} \\
\hline
${^1S_0}$
 & $I=1$  & 
 {$|N \Xi \rangle$} & {$= -\sqrt{\frac{3}{5}} | 8_s \rangle + \sqrt{\frac{2}{5}} | 27 \rangle$} \\
 & & 
 {$|\Sigma \Sigma \rangle$} & {$= \phantom{-}\sqrt{\frac{2}{5}} | 8_s \rangle - \sqrt{\frac{3}{5}} | 27 \rangle$} \\
\hline
${^1S_0}$
 & $I=2$  & 
 {$|\Sigma \Sigma \rangle$} & {$= | 27 \rangle$} \\
\hline \hline
${^3S_1}-{^3D_1}$
 & $I=0$  & 
 {$|N \Xi \rangle$} & {$= | 8_a \rangle$} \\ 
\hline
${^3S_1}-{^3D_1}$
 & $I=1$  & 
 {$|N \Xi \rangle$} & {$= \sqrt{\frac{1}{3}} | 8_a \rangle - \sqrt{\frac{1}{3}} | 10 \rangle - \sqrt{\frac{1}{3}} | \bar{10} \rangle$} \\ 
 & & 
 {$|\Lambda \Sigma \rangle$} & {$= \hspace*{3.08em} + \sqrt{\frac{1}{2}} | 10 \rangle - \sqrt{\frac{1}{2}} | \bar{10} \rangle$} \\
 & & 
 {$|\Sigma \Sigma \rangle$} & {$= \sqrt{\frac{2}{3}} | 8_a \rangle + \sqrt{\frac{1}{6}} | 10 \rangle + \sqrt{\frac{1}{6}} | \bar{10} \rangle$} \\
\hline \hline
\end{tabular*}
\end{center}
\end{table}

\begin{figure}
\begin{tabular}{cc}
 \begin{minipage}[c]{0.5\hsize}
  \includegraphics[scale=0.55]{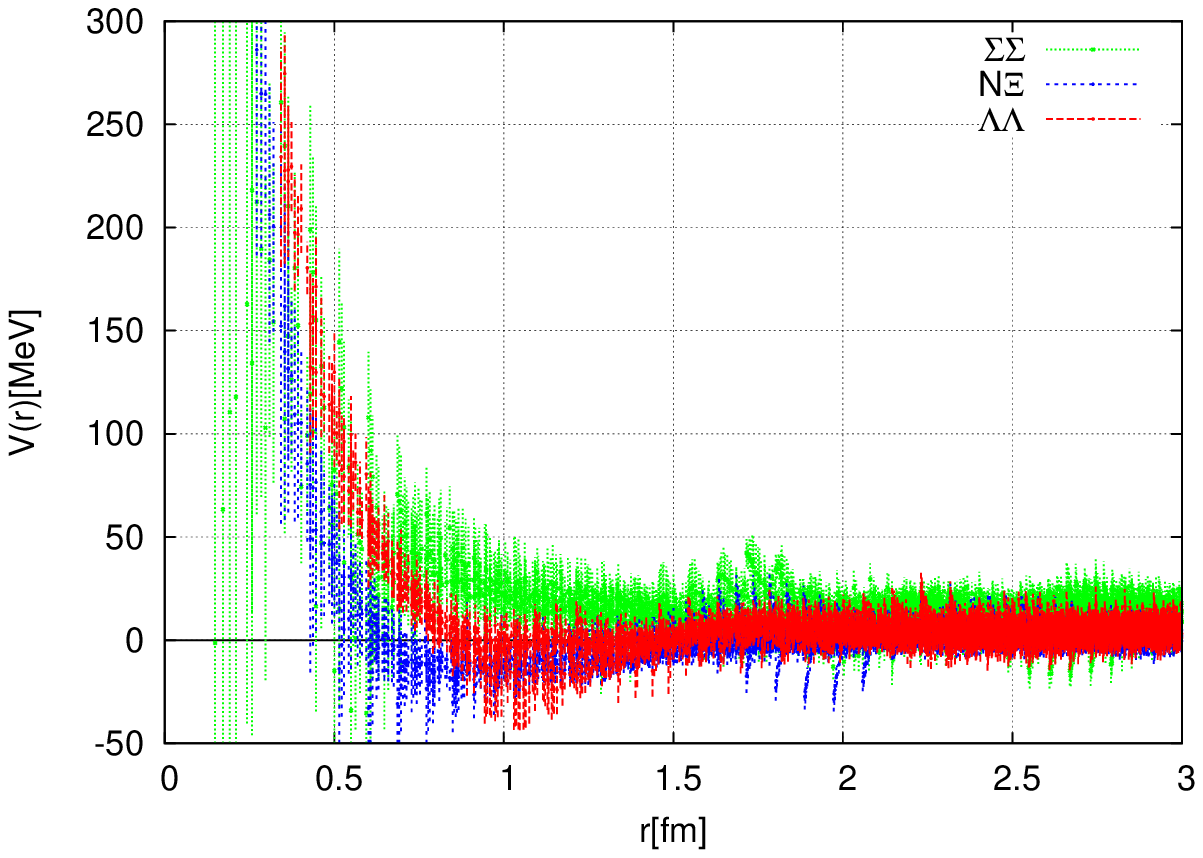}
\end{minipage}
&
\begin{minipage}[c]{0.5\hsize}
  \includegraphics[scale=0.55]{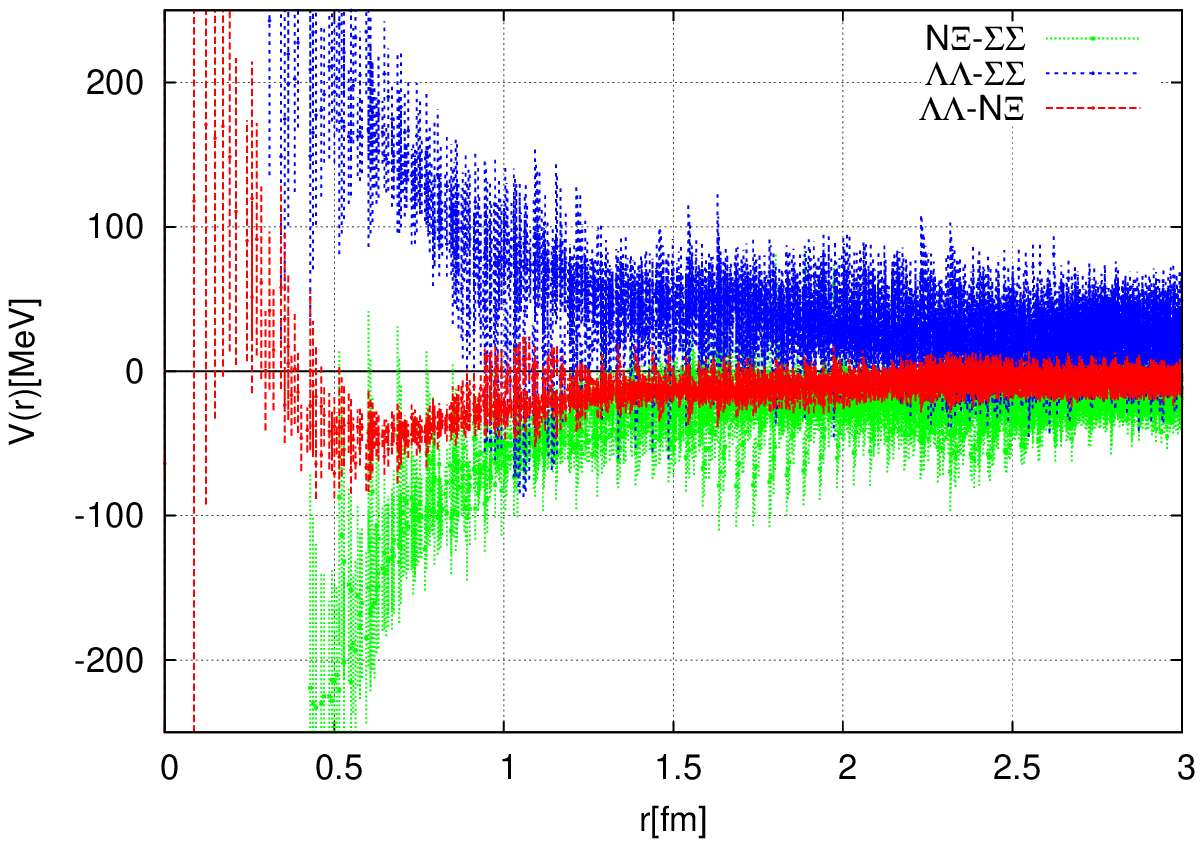}
\end{minipage}
\end{tabular}
\caption{
The $\Lambda \Lambda$(red), $N \Xi$(blue) and $\Sigma \Sigma$(green) potentials are shown in the left panel.
The $\Lambda \Lambda$-$N \Xi$(red), $\Lambda \Lambda$-$N \Xi$(blue) and $N \Xi$-$\Sigma \Sigma$(green) transition potentials are shown in the right panel.}
\label{FIG:1S0I0}
\end{figure}

Fig.~\ref{FIG:1S0I0} shows the coupled channel potential of $\Lambda \Lambda$, $N \Xi$ and $\Sigma \Sigma$ state in the $^1S_0$ ($I=0$) channel, where the existence of $H$ dibaryon state has been discussed as the remaining of bound state in flavor singlet channel \cite{Inoue:2011ai}. 
In the left panel, we find that all potentials of $\Lambda \Lambda$, $N \Xi$ and $\Sigma \Sigma$ have a repulsive core at short distances, while their strength strongly depends on the channel of interest.
A shallow attractive pocket can be seen in ${V^{\Lambda \Lambda}}_{\Lambda \Lambda}$ and ${V^{N \Xi}}_{N \Xi}$, though ${V^{\Sigma \Sigma}}_{\Sigma \Sigma}$ is strongly repulsive even in the long distances.
These features are roughly understandable by the combinations of the potentials in the flavor $SU(3)$ symmetric limit~\cite{Inoue:2011ai}, though our calculations take into account the flavor $SU(3)$ breaking effects.
The ${V^{\Sigma \Sigma}}_{\Sigma \Sigma}$ potential is mainly composed of the flavor octet potential in the $SU(3)$ limit which is known as the forbidden state in non-relativistic quark model and almost kill the attractive contributions from the flavor singlet potential.
On the other hand, the attractive ${V^{N \Xi}}_{N \Xi}$ potential is constructed from large contributions of flavor singlet potential against the small fraction of flavor octet one.
In the right panel of Fig.~\ref{FIG:1S0I0}, the weak ${\Lambda \Lambda}-{N \Xi}$ transition potential comparing other two, ${\Lambda \Lambda}-{\Sigma \Sigma}$ and ${N \Xi}-{\Sigma \Sigma}$, is confirmed.
It suggests that the decay from $N \Xi$ to $\Lambda \Lambda$ is suppressed because of small ${\Lambda \Lambda}-{N \Xi}$ potential and large mass differences of $\Sigma\Sigma$ state from the other two even though their transition potential is large.

\begin{figure}
\begin{tabular}{cc}
 \begin{minipage}[c]{0.5\hsize}
  \includegraphics[scale=0.55]{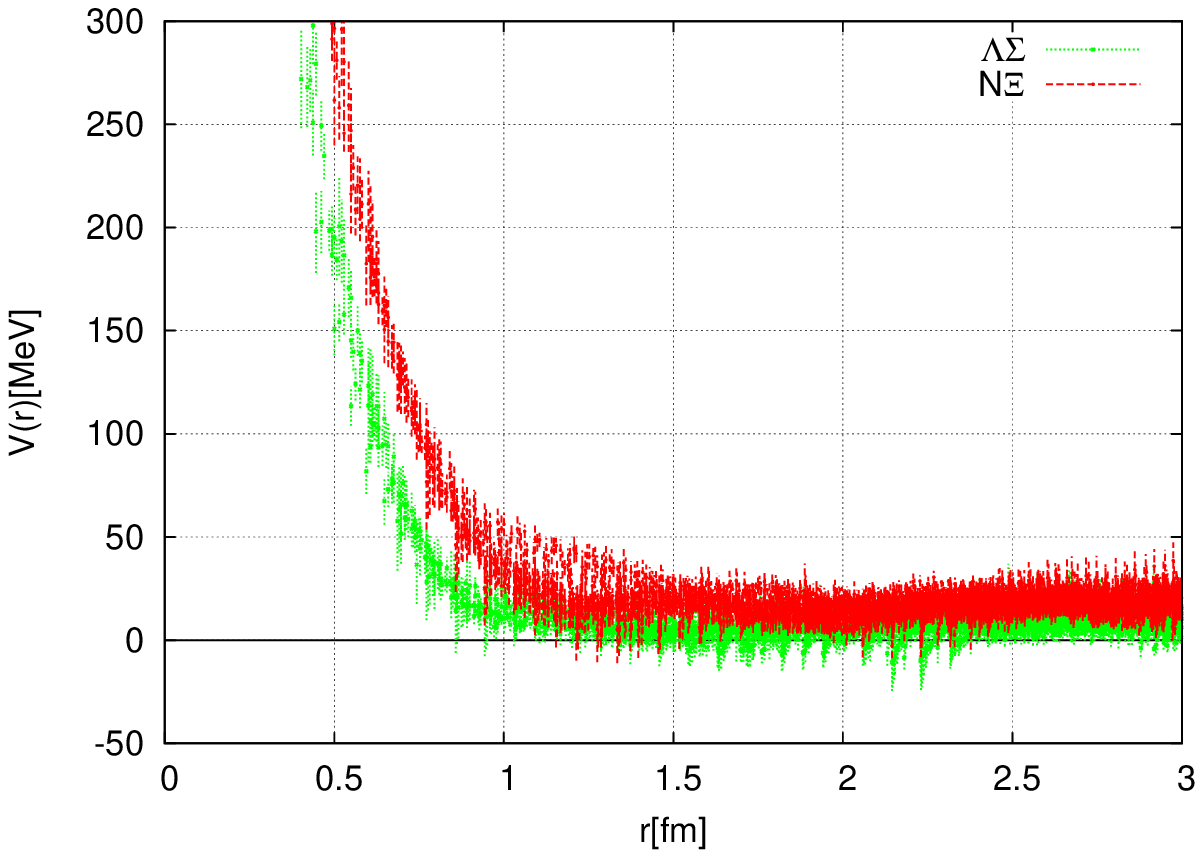}
\end{minipage}
&
\begin{minipage}[c]{0.5\hsize}
  \includegraphics[scale=0.55]{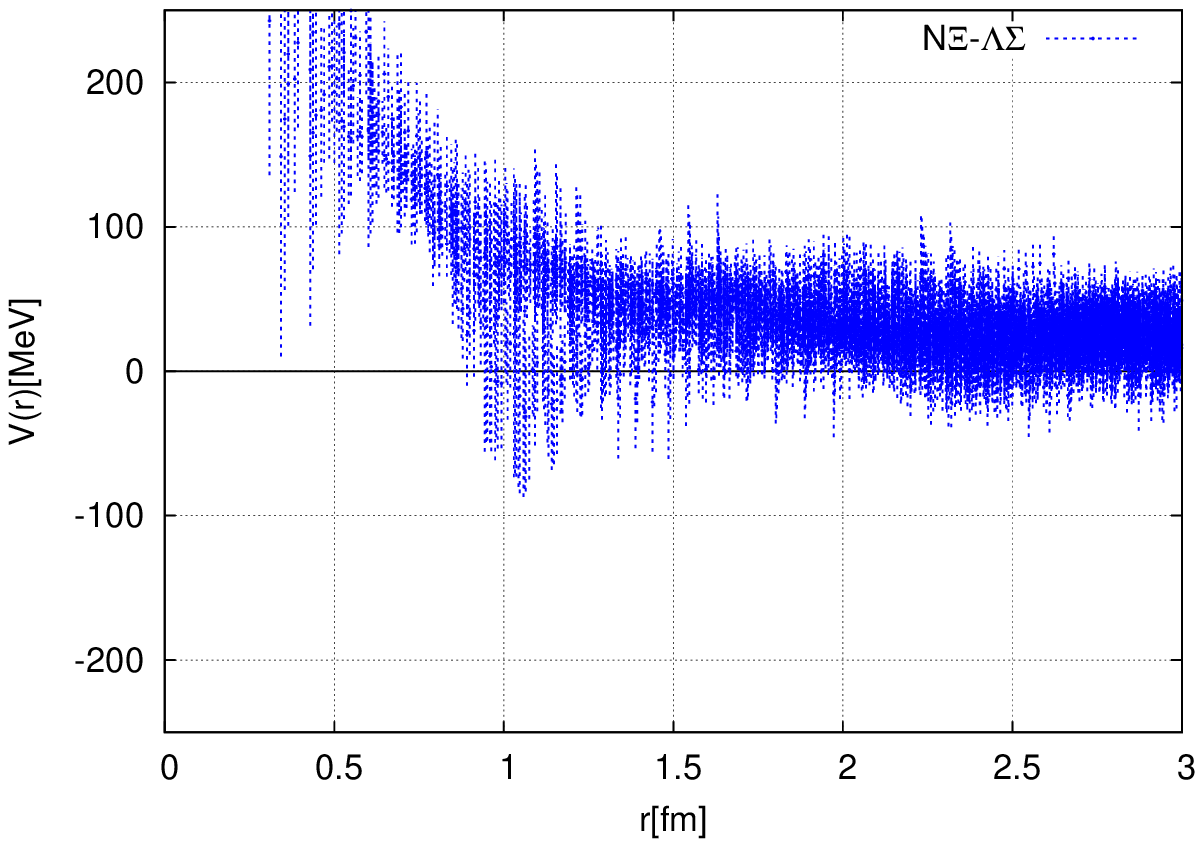}
\end{minipage}
\end{tabular}
\caption{Diagonal (left) and off-diagonal (right) elements of the potential matrix in the ${^1S_0} ( I=1)$ channel.}
\label{FIG:1S0I1}
\end{figure}

A coupled channel potential in ${^1S_0}$ ($ I=1$), which is composed of $N \Xi$ and $\Lambda \Sigma$ states, is given in Fig.~\ref{FIG:1S0I1}.
We find that both the ${V^{N \Xi}}_{N \Xi}$ and ${V^{\Lambda \Sigma}}_{\Lambda \Sigma}$ are repulsive and the strength of $N\Xi$ repulsion is stronger than the $\Lambda \Sigma$ one.
This is again explained by the large contributions of flavor octet potential and the lack of attractive contributions.
By comparing this $N\Xi$ potential with that in $ I=0$ channel, we find that the $N \Xi$ potential has strong dependence on their isospin.
The off-diagonal potential, ${V^{N\Xi}}_{\Lambda\Sigma}$, is comparable to the diagonal elements of potential matrix.
It indicates that the coupling effect between these two states would be important.

\begin{figure}
\begin{tabular}{cc}
\begin{minipage}[c]{0.5\hsize}
  \includegraphics[scale=0.55]{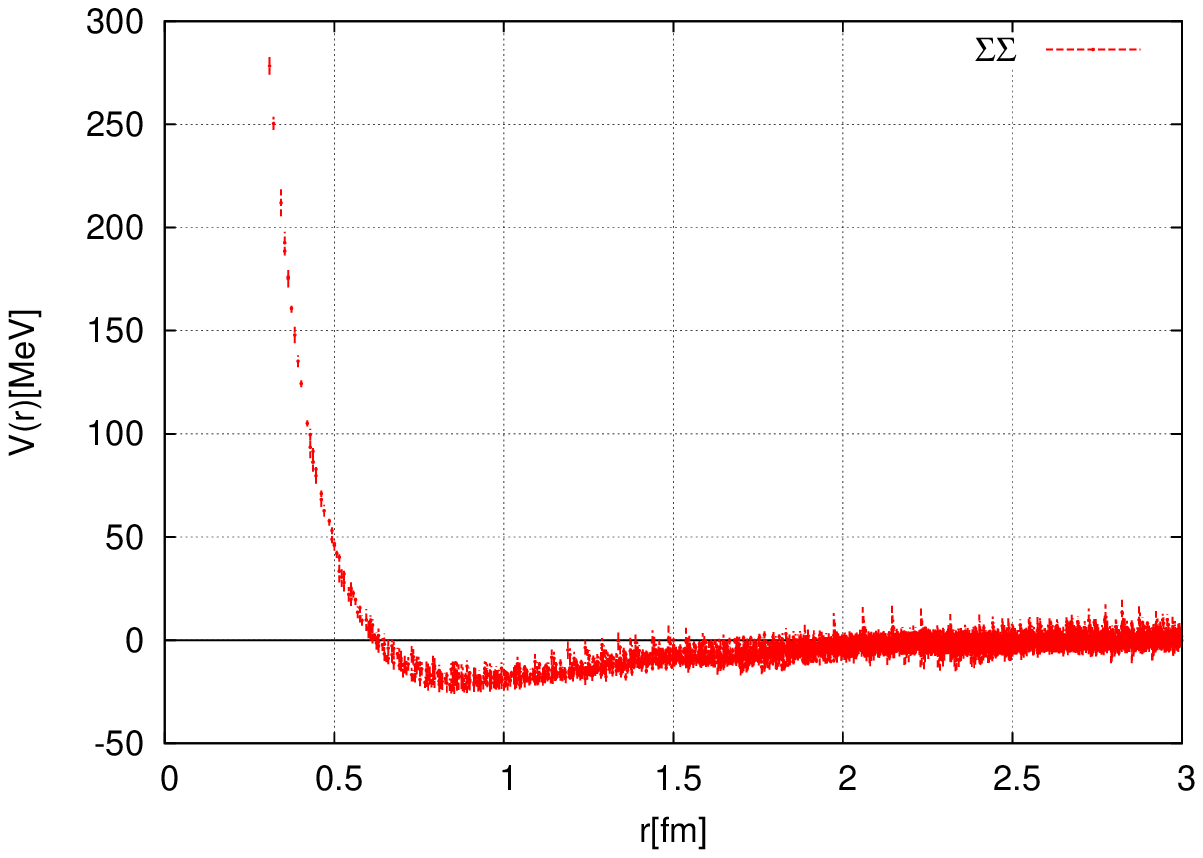}
\end{minipage} &  
\begin{minipage}[c]{0.5\hsize}
  \includegraphics[scale=0.55]{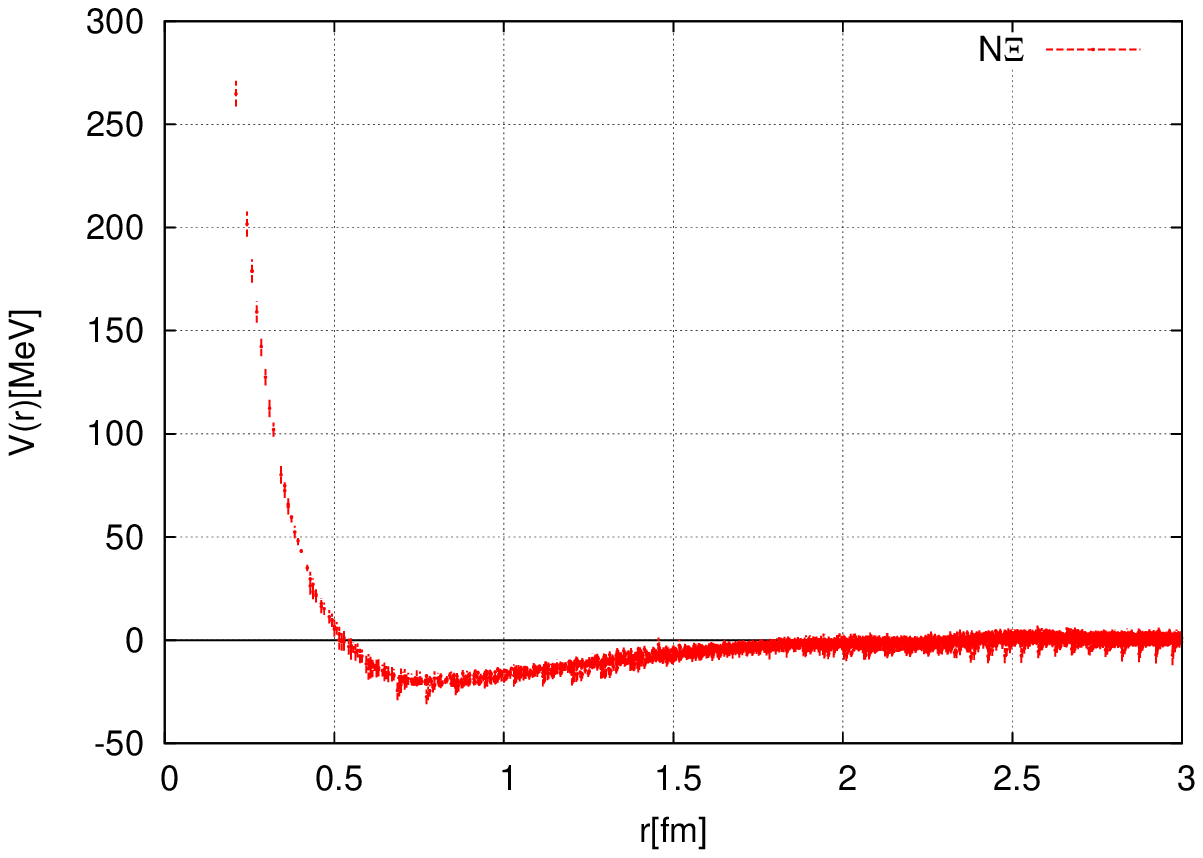}
\end{minipage}
\end{tabular}
\caption{The $\Sigma\Sigma$ potential in the ${^1S_0} (I=2)$ channel (left) and the $N\Xi$ potential in the ${^3S_1} (I=0)$ channel  (right). }
\label{FIG:1S0I2}
\end{figure}

Fig.~\ref{FIG:1S0I2} shows the $\Sigma\Sigma$ potential in ${^1S_0} (I=2)$ channel (left) and the $N\Xi$ potential in the ${^3S_1} (I=0)$ channel (right).
The $\Sigma\Sigma$ potential in the ${^1S_0} (I=2)$ channel (left), which belongs to the $\boldmath{27}$-plet irreducible representation in the flavor $SU(3)$, has repulsion at short distance and attraction at long distance.
Similarly, the $N\Xi$ potential in the  ${^3S_1} (I=0)$ channel (right) has
a repulsion at short distance,  which is surrounded by a long range attraction. 
Despite the qualitative similarity between two potentials, the repulsion of the $N\Xi$ potential at short distance is much weaker than that of the $\Sigma\Sigma$ potential. 
This difference may be interpreted to the Pauli blocking effect among the constituent quarks.

\begin{figure}
\begin{tabular}{cc}
 \begin{minipage}[c]{0.5\hsize}
  \includegraphics[scale=0.55]{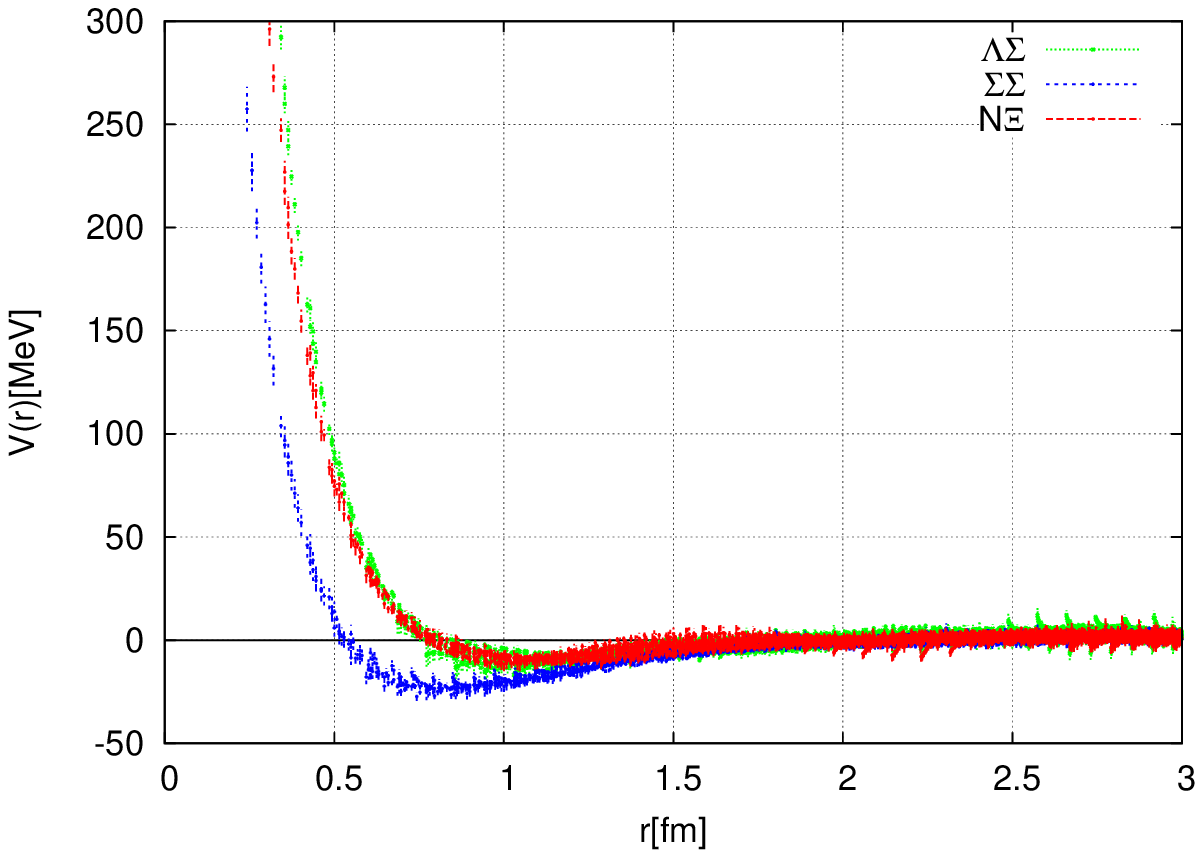}
\end{minipage}
&
\begin{minipage}[c]{0.5\hsize}
  \includegraphics[scale=0.55]{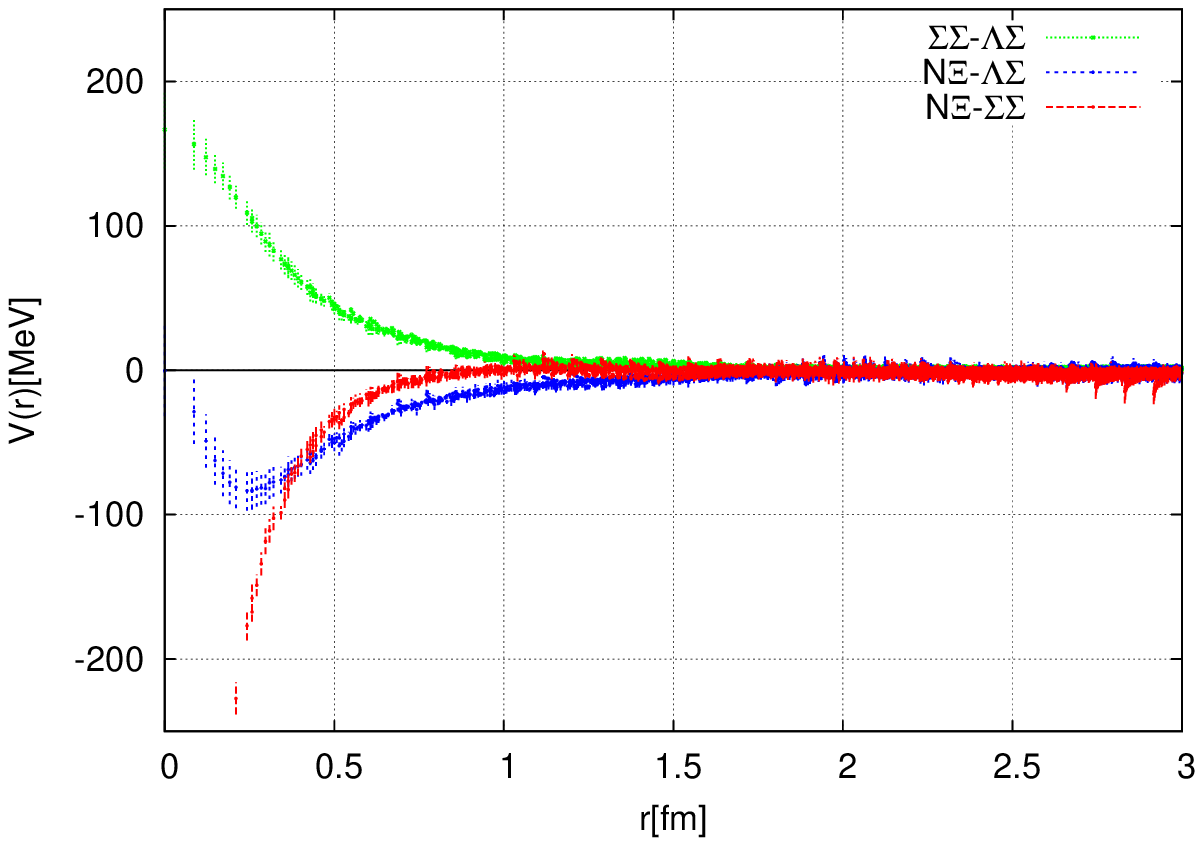}
\end{minipage}
\end{tabular}
\caption{
Diagonal (left) and off-diagonal (right) elements of the potential matrix in the ${^3S_1} (I=1)$ channel.
}
\label{FIG:3S1I1}
\end{figure}

As it can be seen in Fig.~\ref{FIG:3S1I1} where potentials in  the ${^3S_1}$ ($I=1$) channel, which has $N \Xi$, $\Lambda \Sigma$ and $\Sigma \Sigma$ components, are shown, we find an attraction enclosing a repulsive core at short distance for all potentials of $N\Xi$, $\Lambda\Sigma$ and $\Sigma\Sigma$ state.
Among them the largest attraction comes from ${V^{\Sigma \Sigma}}_{\Sigma\Sigma}$.
For the off-diagonal potentials, all transition potentials, ${V^{N\Xi}}_{\Lambda \Sigma}$, ${V^{\Lambda \Sigma}}_{\Sigma \Sigma}$ and ${V^{N \Xi}}_{\Sigma \Sigma}$, are very small in $r > 1.0$ fm.
These results indicate that the coupled channel effect could be small in this channel.
When we check the isospin dependence of the spin triplet $N \Xi$ potentials in ${^3S_1}$, we find that the $I=0$ potential has stronger attraction and weaker repulsive core than that with $I=1$.

\section{Conclusions}
We have investigated $S=-2$ $BB$ interactions from lattice QCD employing $N_f = 2+1$ gauge configurations with $(96a)^4$ and $a \simeq 0.085$fm lattice, where $m_\pi \simeq 146$ MeV and $m_K \simeq 525$ MeV.
Baryon potentials have been calculated by the coupled channel HAL QCD method with considerations of not only spacial but also temporal correlations of baryon four-point correlation functions which involves the equal-time NBS wave functions.

We find that all diagonal elements of potential matrix in $S=-2$ $BB$ system have a short-range repulsive core but
their strengths are strongly depend on their flavors (isospins) and spins, which might be understood with a help of $SU(3)$ CG coefficients and a consequence of Pauli blocking effect of consistent quarks.
From the off-diagonal potential in ${^1S_0}$ $(I=0)$ channel, we expect that the strong decay of $N\Xi$ in  to $\Lambda \Lambda$, which is important for the lifetime of $\Xi$-hypernuclei, is suppressed.
We also find that the $N \Xi$ potential with $I=0$ is more attractive than that with $I=1$ for both ${^1S_0}$ and ${^3S_1}$ cases.
We expect that the spin and isospin dependence of $N\Xi$ interaction will be confirmed by relativistic heavy-ion collisions which provide an interesting experimental opportunities to study baryon forces.

The results in this paper are still very preliminary but further investigations will be performed with high statistics data.

\section*{Acknowledgments}
We thank members of PACS Collaboration for the gauge configuration generation. The lattice QCD calculations have been performed on the K computer at RIKEN, AICS (hp120281, hp130023, hp140209, hp150223, hp150262, hp160211), HOKUSAI FX100 computer at RIKEN, Wako (G15023, G16030) and HA-PACS at University of Tsukuba (14a-20, 15a-30). We thank ILDG/JLDG~\cite{JLDGILDG} which serves as an essential infrastructure in this study. This work is supported in part by MEXT Grant-in-Aid for Scientific Research (JP15K17667), SPIRE (Strategic Program
for Innovative REsearch) Field 5 project and "Priority Issue on Post-K computer" (Elucidation of the Fundamental Laws and Evolution of the Universe).



\end{document}